\documentclass[aps,prd,twocolumn,nofootinbib,showpacs]{revtex4-1}
\usepackage{amsmath}
\usepackage{amssymb}
\usepackage{dcolumn}
\usepackage{graphicx}
\usepackage{verbatim}
\usepackage{subfigure}
\usepackage[colorlinks, linkcolor=blue, citecolor=blue, urlcolor=blue]{hyperref}

\newcommand{\f}{\frac}
\newcommand{\ep}{\epsilon}
\newcommand{\om}{\omega}
\newcommand{\ph}{\phi_{\scriptscriptstyle\rm VEV}}
\newcommand{\Mpl}{M_{\mathrm{Pl}}}

\def\prl{Phys. Rev. Lett.}
\def\prd{Phys. Rev. D}
\def\aap{Astron. Astrophys.}
\def\apjl{Astrophys. J. Lett.}
\def\apj{Astrophys. J.}
\def\jcap{JCAP}
\def\mnras{Mon. Not. Roy. Astron. Soc.}
\def\sovast{Soviet Ast.}

\begin{document}

\title{Constraining the scalar-tensor gravity theories with and without screening mechanisms by combined observations}

\author{Xing Zhang$^{1,2}$}\email{starzhx@ustc.edu.cn}
\author{Rui Niu$^{1,2}$}
\author{Wen Zhao$^{1,2}$}\email{wzhao7@ustc.edu.cn}

\affiliation{$^1$ CAS Key Laboratory for Researches in Galaxies and Cosmology, Department of Astronomy, \\ University of Science and Technology of China, Chinese Academy of Sciences, Hefei, Anhui 230026, China}
\affiliation{$^2$ School of Astronomy and Space Science, University of Science and Technology of China, Hefei 230026, China}

\date{\today}

\begin{abstract}
Screened modified gravity (SMG) and Brans-Dicke (BD) gravity are typical examples of scalar-tensor theories with and without screening mechanisms, which can suppress the scalar field in dense regions.
In this paper, we investigate the tests of time-varying gravitational constant $G$, gravitational dipole radiation, and Nordtvedt effect in BD and SMG theories, respectively.
We place new constraints on these theories by combining Cassini experiment, lunar laser ranging (LLR) measurements, and pulsar observations from PSRs J1738$+$0333 and J0348$+$0432.
We find that screening mechanism has important influence on theoretical constraints.
The strongest, second, and weakest constraints on BD are from Cassini, pulsar, and LLR tests, respectively.
The most stringent constraint on SMG comes from LLR measurements and improves the previous best constraint by more than seven orders of magnitude.
We derive the bounds on the cosmological evolution of the scalar background in these theories using the time variation of $G$.
The results of all tests agree well with general relativity (GR) and give more stringent constraints on the deviations from GR.
Finally, as an example, we consider the chameleon model and derive the constraints on the model parameters.
\end{abstract}

\maketitle

\section{Introduction}\label{section1}
Einstein's general relativity (GR) is one of the two pillars in modern physics.
Nevertheless, it suffers from the dark matter and dark energy problems \cite{Cline:2013aa,Sahni:2004ai}, and cannot be quantized as other field theories \cite{Kiefer:2007aa,DeWitt:1967yk}.
Therefore, there are countless attempts to develop alternative theories of gravity. 
One of the most simple and popular alternative theories is scalar-tensor theory \cite{Fujii:2003ab,Damour:1992ab,Gannouji:2006aa}, in which the gravitational interaction is mediated by an underlying scalar field and the tensor field of GR.

In scalar-tensor theory, the gravitational constant $G$ is controlled by the background scalar field which can vary with the expansion of the Universe \cite{Uzan:2011aa}. 
The time variation of $G$ generally indicates a violation of strong equivalence principle (SEP) \cite{Will:1993aa,Will:2014aa}, and leads to an important contribution to the time change in the orbital period of the binary systems \cite{Nordtvedt:1990aa}.
The SEP violation causes the two centers of gravitational and inertial masses of compact system to separate from each other, which induces a mass dipole moment.
As binaries orbit each other, the mass dipole moment will emit gravitational dipole radiation, which dominates the orbital decay of asymmetric binary systems.
Therefore, one often has to consider both the variation of $G$ and the dipole radiation in a strong-field testing gravity with binary pulsars \cite{Nice:2005aa, Lazaridis:2009aa, Freire:2012aa, Zhu:2015aa, Zhu:2018aa}.
Moreover, because of the SEP violation, compact objects with different gravitational self-energy feel different accelerations in the additional field, which is called the Nordtvedt effect \cite{Nordtvedt:1990aa} and has been tightly constrained by lunar laser ranging (LLR) experiment \cite{Hofmann:2010aa}.

Both Brans-Dicke (BD) gravity \cite{Brans:1961aa} and screened modified gravity (SMG) \cite{Brax:2012aa} are examples of a scalar-tensor theory.
BD gravity is the earliest and most widely studied scalar-tensor theory of gravity, and it predicts stronger non-GR effects in strong-field regime \cite{Will:1993aa,Will:2014aa}.
Therefore, the strong-field tests of scalar-tensor gravity are mostly based on BD gravity.
SMG is a kind of scalar-tensor theories of gravity with screening mechanisms, including chameleon \cite{Khoury:2004aa}, symmetron \cite{Hinterbichler:2010es}, dilaton \cite{Damour:1994ab}, and $f(R)$ \cite{Sotiriou:2010aa,De-Felice:2010aa} theories.
SMG theories operate an environment-dependent scalar field and suppress the scalar force in dense regions, therefore the deviations from GR become smaller and weaker in strong-field regime, which is completely different from BD gravity (without screening mechanism).

In this paper, we study the gravitational dipole radiation, the time variation of $G$, and the Nordtvedt effect in BD and SMG theories, respectively, which allows us to test the violation of SEP with current observations.
We place new constraints on dipole radiation and time-varying $G$ by performing Monte Carlo simulations and combining the Solar system tests (LLR measurements and Cassini experiment) and the pulsar observations (from PSRs J1738$+$0333 and J0348$+$0432).
It turns out that there are huge differences between the constraints on theories with and without screening mechanisms. 
The Cassini experiment, pulsar observations, and LLR measurements are the strongest, second, and weakest constraints on BD gravity, respectively.
We find that the most stringent constraint on SMG from LLR measurements in the Earth-Moon system is several order of magnitude more stringent than the pulsar observations, because in this theory the binary pulsar is a strong screening system relative to the Earth-Moon system.
The LLR constraint on $\dot{G}$ in the two theories is also more stringent than the pulsar observations.
Using the constraints on $\dot{G}$, we derive the constraints on the time evolution of the scalar background in the two theories, which differ by several orders of magnitude because of screening mechanism.
The results about SMG are generically applicable, as an example, we consider the exponential chameleon model and derive the constraints on this model.
In this paper all tests show good agreement with GR and all the previous works and yield very stringent constraints on the strong-field deviations from GR.

The plan of the paper is as follows.
In Sec. \ref{section2}, we study the orbital period decay effect and the Nordtvedt effect in SMG and BD gravities, respectively.
In Sec. \ref{section3}, we place the constraints on the two theories by combining the observations of the Solar system and the binary pulsars.
In Sec. \ref{section4}, we apply our results to the exponential chameleon model and derive the constrains on the model parameters.
Our conclusions are summarised in Sec. \ref{section5}.

\section{BD and SMG}\label{section2}
In this section we investigate the orbital period decay of binary pulsars caused by dipole radiation and time-varying $G$ in BD and SMG theories, then discuss the Nordtvedt effect due to the violation of SEP in these theories. 
These will allow us to place constraints on these theories.

\subsection{Orbital Period Decay}
The change in the orbital period of the binary pulsars is related to the damping of the orbital energy due to the emission of gravitational waves (GWs).
In fact, the monitoring of the orbital period led to the first indirect detection of GWs \cite{Hulse:1975aa,Taylor:1982aa,Taylor:1989aa}.
The orbital period decay caused by dipole radiation in an elliptical binary system is given by \cite{Will:1993aa,Will:2014aa,Zhang:2019ab,Zhang:2019aa}
\begin{align}
\label{P_dipole}
\dot{P}_b^{\rm D}=-\f{4\pi^2G}{P_b}\f{m_1m_2}{m_1+m_2}\kappa_{\rm D}(s_1-s_2)^2
                     \f{1+e^2/2}{(1-e^2)^{5/2}},
\end{align}
where $\kappa_{\rm D}$ is a model-dependent constant that quantifies the contribution of the dipole radiation induced by self-gravity, $\kappa_{\rm D}=0$ in GR, but it is generally not the case in SEP-violating theories of gravity.
The quantity $s_i$ is the $i$-th object's sensitivity, defined by $s_i\equiv({d\ln m_i}/{d\ln \phi})_0$ \cite{Eardley:1975aa}, which describes the response of the gravitational binding energy to the external gravitational field.
In general, $\kappa_{\rm D}$ and $s_i$ take different values for different theories of gravity.
In these theories under consideration, $\kappa_{\rm D}=2/(\om_{\rm BD}+2)$ and $s_i=E_i^{\rm g}$ for BD \cite{Will:1993aa,Will:2014aa}, and $\kappa_{\rm D}=2\Mpl^2/\ph^2$ and $s_i=\ph^2/(2\Mpl^2\Phi_i)$ for SMG \cite{Zhang:2019ab}. 
Here, $\om_{\rm BD}$ is a dimensionless parameter of BD gravity, and $\ph$ is the scalar background\footnote{In this paper, the scalar background in BD is labeled as $\phi_0$.} in SMG (i.e., the vacuum expectation value (VEV) of the scalar field).
The quantity $E_i^{\rm g}$ ($E_i^{\rm g}\ge0$) is the negative of the gravitational self-energy per unit mass, $\Phi_i=Gm_i/R_i$ is the $i$-th object's compactness (i.e., negative gravitational potential at the surface), and they are of the same order of magnitude.
Note that $s_i$ in BD is positively correlated to the compactness, but on the contrary in SMG, $s_i$ is inversely proportional to the compactness, which is the most important difference between the theories with and without screening mechanisms.

In scalar-tensor theory, the gravitational constant $G$ can become time-dependent and vary with the expansion of the Universe. 
Nordtvedt \cite{Nordtvedt:1990aa} first pointed out that a time-varying $G$ leads to an additional contribution to the change in the orbital period of the binary system.
To leading order, the change rate in the orbital period due to $\dot{G}$ is given by \cite{Nordtvedt:1990aa}
\begin{align}
\label{Pbdot_Gdot}
\dot{P}_b^{\dot{G}}=-2P_{b}\f{\dot{G}}{G}\bigg[1+\f{2m_1+3m_2}{2(m_1+m_2)}c_1
                             +\f{3m_1+2m_2}{2(m_1+m_2)}c_2\bigg],
\end{align}
where $c_i$ is the body-dependent quantity, defined by $c_i\equiv({d\ln m_i}/{d\ln G})_0$ \cite{Nordtvedt:1990aa}.
In these theories under consideration, $c_i=-s_i$ for $G\propto1/\phi_0$ in BD \cite{Will:1993aa,Will:2014aa}, and $c_i=\Phi_1\Phi_2/(2\Phi_i)$ for $G\propto1+\ph^2/(2\Mpl^2\Phi_1\Phi_2)$ in SMG \cite{Zhang:2017aa}.

In most alternative theories of gravity, both dipole radiation and time-varying $G$ appear simultaneously. 
Therefore, one often has to consider both effects $\dot{P}_b^{\rm D}+\dot{P}_b^{\dot{G}}$ for testing GR using binary pulsars.

\subsection{Nordtvedt Effect}
Most alternative theories of gravity predict that strongly self-gravitating bodies do not follow geodesics of the background spacetime, massive bodies with different gravitational binding energy feel different accelerations, which leads to the violation of SEP in these theories.
This is known as the Nordtvedt effect \cite{Nordtvedt:1968aa,Nordtvedt:1968ab}, usually parametrized by the Nordtvedt parameter $\eta_\mathrm{N}$ (also called the SEP violation parameter),
\begin{align}
\f{m_{\rm g}}{m_{\rm i}}=1-\eta_{\rm N}E^{\rm g},
\end{align}
where $E^{\rm g}$ is the negative of the gravitational self-energy per unit mass, and $m_{\rm i}$ and $m_{\rm g}$ are the inertial and gravitational masses of bodies, respectively.
In GR $\eta_{\rm N}=0$, but it is generally nonzero in alternative theories that violate the SEP.
The relative acceleration of two bodies $a$ and $b$ in an external gravitational potential $U_c$ is then
\begin{align}\label{acceler_ab2}
\mathbf{a}_{ab}=\eta_{\rm N}(E^{\rm g}_a-E^{\rm g}_b)\nabla{U_c},
\end{align}
where $U_c=-Gm_c/r$ for a source of spherically symmetric mass distribution.
This leads to detectable effects in the Solar system. 
Most notably, it leads to an Earth-Moon range oscillation in the gravitational field of the Sun \cite{Nordtvedt:1968ab, Nordtvedt:1982aa}, which can be constrained by using LLR experiment \cite{Hofmann:2010aa}.

In BD gravity, $\eta_{\rm N}$ is a body-independent constant, given by $\eta_{\rm N}=1/(\om_{\rm BD}+2)$ \cite{Will:1993aa,Will:2014aa}.
Below let us derive the Nordtvedt parameter in SMG.
The Nordtvedt effect results in an anomalous difference in the accelerations of two different objects in an external gravitational field.
Therefore, the Nordtvedt parameter can be extracted from the equations of motion.
In the previous work \cite{Zhang:2019ab}, we have derived in detail the $n$-body equations of motion in SMG, and up to Newtonian order, given by
\begin{align}
\begin{split}
\label{N_body_eom}
\mathbf{a}_{a}&=-\sum_{b\ne a}\frac{{\mathcal G}_{ab}m_b}{r^2_{ab}}\mathbf{\hat{r}}_{ab},
\end{split}
\end{align}
with
\begin{align}\label{G_ab2}
{\mathcal G}_{ab}= G\Big(1+\frac12\ep_a\ep_b\Big),
\end{align}
where $\mathbf{a}_{a}\equiv d^2\mathbf{r}_{a}/dt^2$ is the acceleration of the $a$-th object, $\mathbf{\hat{r}}_{ab}$ is the unit direction vector from the $b$-th object to the $a$-th object, and $r_{ab}=\left|\mathbf{r}_a-\mathbf{r}_b\right|$. 
The quantity ${\mathcal G}_{ab}$ is the effective gravitational constant between two objects $a$ and $b$, $\ep_i$ is the scalar charge of the $i$-th object, defined by $\ep_i=\ph/(\Mpl\Phi_i)$ \cite{Zhang:2019ab}.
Now, considering a pair of bodies $a$ and $b$ moving in the gravitational field of a third body $c$, in the case of $r_{ab}\ll r_{ac}$ and $r_{ac}\simeq r_{bc}$, from Eq. \eqref{N_body_eom} the relative acceleration is
\begin{align}\label{acceler_ab1}
\mathbf{a}_{ab}&=-\f{1}{2}(\ep_a-\ep_b)\ep_c\frac{Gm_c}{r^2_{ac}}\mathbf{\hat{r}}_{ac}.
\end{align}
By comparing with \eqref{acceler_ab2}, yields
\begin{align}
\eta_{\rm N}=\f{\ph^2}{2\Mpl^2}\f{1}{\Phi_a\Phi_b\Phi_c}\f{\Phi_a-\Phi_b}{E^{\rm g}_a-E^{\rm g}_b}.
\end{align}
Note that, unlike $\eta_{\rm N}$ in BD (body-independent constant), $\eta_{\rm N}$ in SMG is the three-body-dependent parameter.
The results of this section will allow us to test these theories of gravity with current observations in the next section.

\section{Constraints}\label{section3}
The tests of dipole radiation and time-varying $G$ need any two binary pulsars with different orbital periods to break the degeneracy between Eqs. \eqref{P_dipole} and \eqref{Pbdot_Gdot}, because of $\dot{P}_b^{\rm D}\propto P_b^{-1}$ and $\dot{P}_b^{\dot{G}}\propto P_b$. 
We use two different pulsar-white dwarf binaries, namely PSRs J1738$+$0333 \cite{Freire:2012aa} and J0348$+$0432 \cite{Antoniadis:2013aa}, with the system parameters shown in Table \ref{tab1_psr}.
Note that the intrinsic value of $\dot{P}_b$ has been given by subtracting the kinematic effect \cite{Damour:1991aa} and Shklovskii effect \cite{Shklovskii:1970aa} from its observed value.
The excess orbital period change $\dot{P}_b^{\rm Exc}$ can be obtained by subtracting $\dot P_b^{\rm GR}$ predicted by GR's quadrupole radiation from the intrinsic $\dot P_b^{\rm Int}$. 
The excess $\dot{P}_b^{\rm Exc}$ comes from the contributions of the below physical effects, $\dot{P}_b^{\rm Exc}=\dot{P}_b^{\dot{M}}+\dot{P}_b^{\rm T}+\dot{P}_b^{\dot{G}}+\dot{P}_b^{\rm D}$ \cite{Damour:1991aa}, where $\dot{P}_b^{\dot{M}}$ and $\dot{P}_b^{\rm T}$ are the contributions from mass loss and tidal effects, respectively.
In the two systems under consideration, $\dot{P}_b^{\dot{M}}$ and $\dot{P}_b^{\rm T}$ can be neglected, because they are much smaller than the uncertainty in the measurement of the excess $\dot{P}_b^{\rm Exc}$.

\begin{table*}[htbp]
\centering
\caption{Parameters of the binary pulsar systems with 1-$\sigma$ uncertainties.\label{tab1_psr}}
\begin{tabular}{lrr}
\hline\hline
 PSR    &    J1738$+$0333 \cite{Freire:2012aa}    &    J0348$+$0432 \cite{Antoniadis:2013aa} \\
 \hline
Eccentricity, $e$ ($10^{-6}$)    &     $0.34\pm0.11$                 &    $2.4\pm1.0$ \\
Orbital period, $P_b$ (day)       &     $0.3547907398724(13)$   &    $0.102424062722(7)$ \\
Intrinsic period derivative, $\dot P_b^{\rm Int}$ ($10^{-13}$ s s$^{-1}$)          &   $-0.259\pm0.032$             &    $-2.73\pm0.45$ \\
$\dot{P}_b$ predicted by GR, $\dot{P}_b^{\rm GR}$ ($10^{-13}$ s s$^{-1}$)   &   $-0.277^{+0.015}_{-0.019}$  &    $-2.58^{+0.08}_{-0.11}$ \\
Pulsar mass, $m_p$ ($M_{\odot}$)          &    $1.46^{+0.06}_{-0.05}$         &    $2.01\pm0.04$ \\
Companion mass, $m_c$ ($M_{\odot}$)   &    $0.181^{+0.008}_{-0.007}$    &    $0.172\pm0.003$ \\
Companion radius, $R_c$ ($R_{\odot}$)   &    $0.037^{+0.004}_{-0.003}$   &    $0.065\pm0.005$ \\
\hline\hline
\end{tabular}
\end{table*}

\begin{figure*}[!htbp]
\centering
\subfigure[BD]{
\includegraphics[width=8cm, height=8cm]{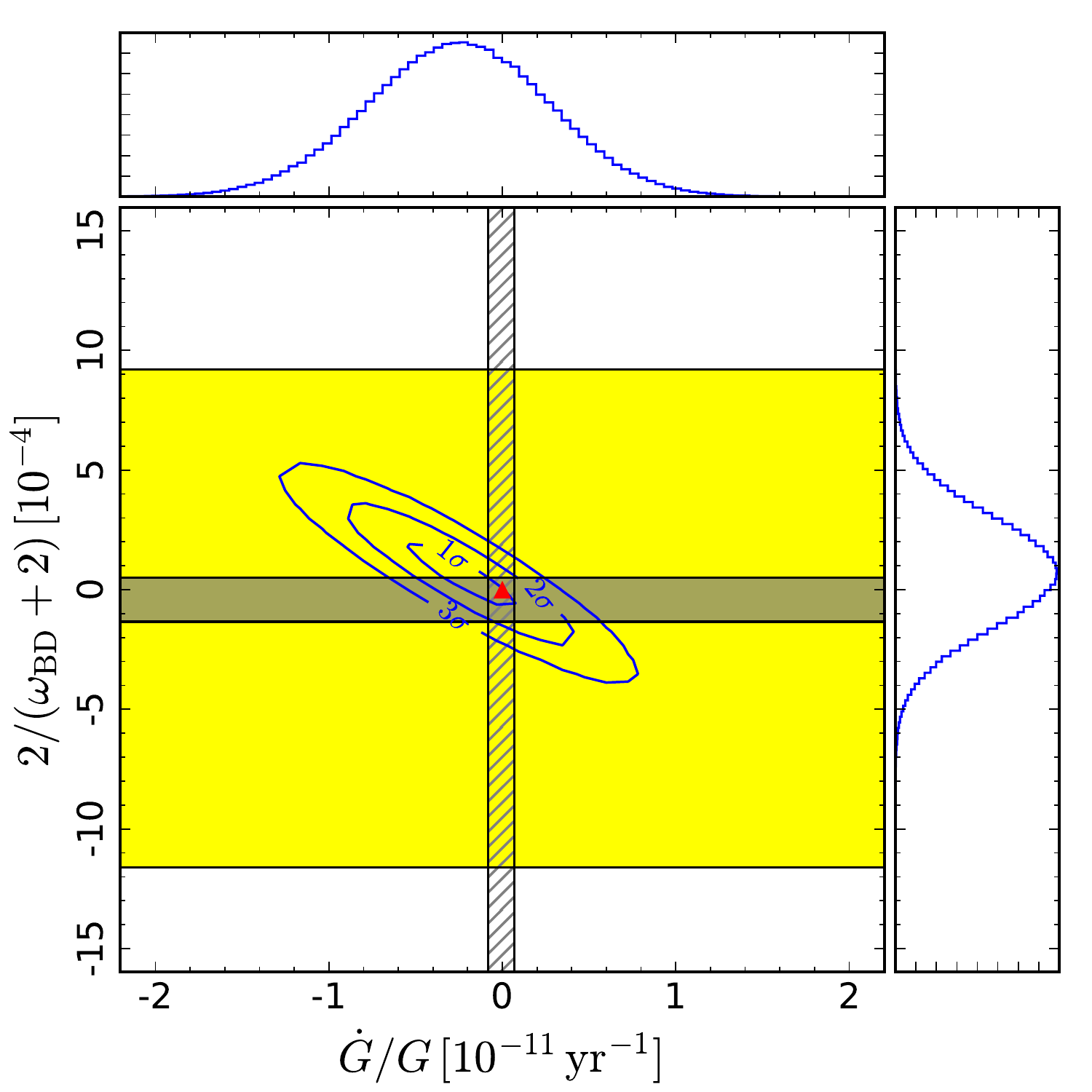}}
\hspace{5mm}
\subfigure[SMG]{
\includegraphics[width=8cm, height=8cm]{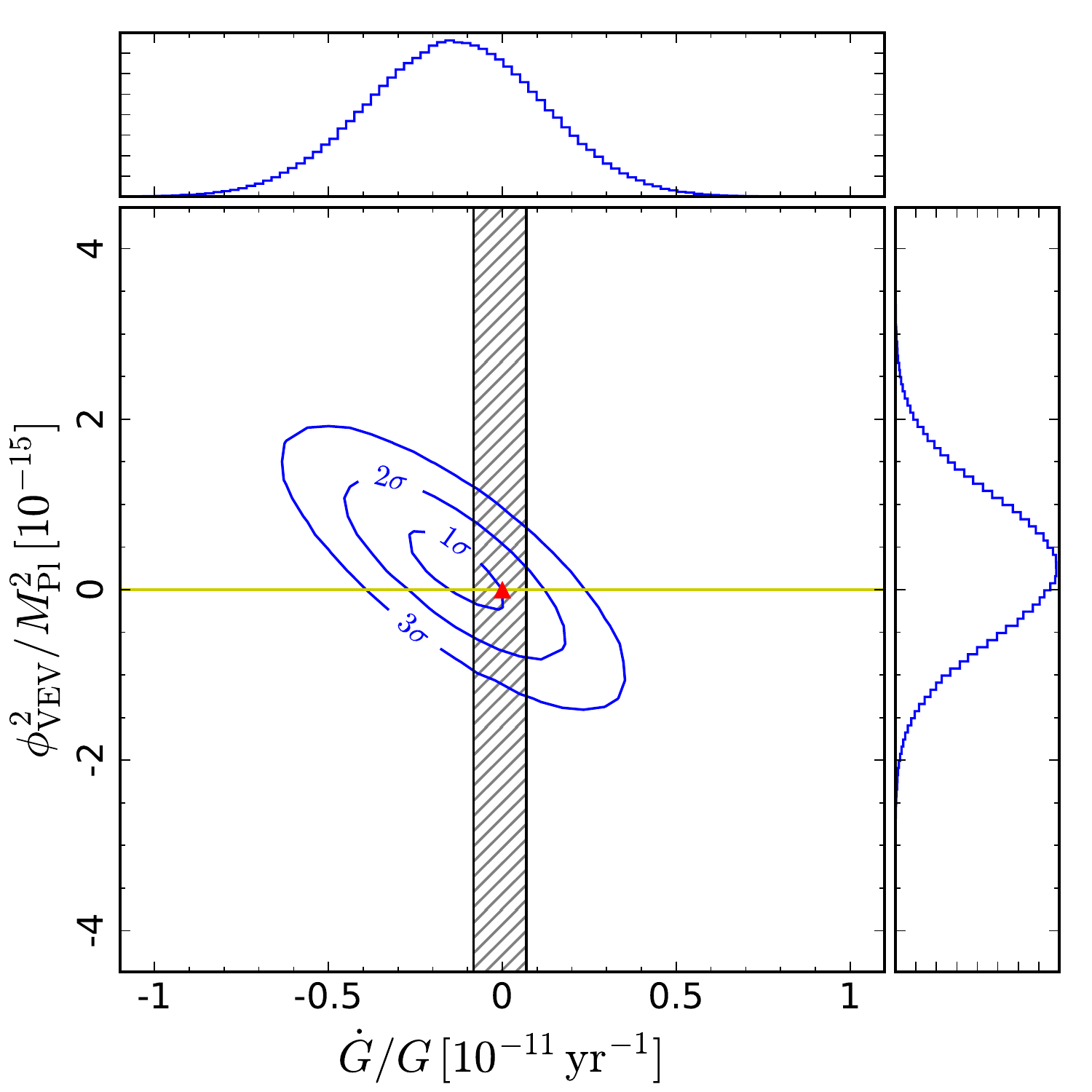}}
\caption{Confidence contours of $\dot{G}/G$ and dipole radiation parameter ($2/(\om_{\rm BD}+2)$ for BD and ${\ph^2}/{\Mpl^2}$ for SMG) based on PSRs J1738$+$0333 and J0348$+$0432 using Monte Carlo simulations. 
The yellow region and shaded region mark the 95.4\% confidence limit from LLR \cite{Hofmann:2010aa}, and the gray region marks the 95.4\% confidence limit from the Cassini experiment \cite{Bertotti:2003aa}. 
The red triangle represents the GR result.}
\label{fig_psr_llr_BD_SMG}
\end{figure*}

Each pulsar system provides a constraint on $\dot{P}_b^{\rm Exc}$, which allows us to test the dipole radiation and time-varying $G$ over the time span of the observation.
Using this and Eqs. \eqref{P_dipole} and \eqref{Pbdot_Gdot}, and performing $10^6$ Monte Carlo simulations, yields the constraints on dipole radiation and time-varying $G$ as shown in Figure \ref{fig_psr_llr_BD_SMG}. 
The individual constraints at 68.3\%~confidence level (CL) are for BD
\begin{align}
\f{\dot{G}}{G}=(-2.5\pm5.2)\times10^{-12}~{\rm yr}^{-1},\label{BD_G_W}
\\
\f{2}{\om_{\rm BD}\!+\!2}=(0.7\pm2.3)\times10^{-4},
\end{align}
and for SMG
\begin{align}
\f{\dot{G}}{G}=(-1.4\pm2.5)\times10^{-12}~{\rm yr}^{-1},
\\
\f{\ph^2}{\Mpl^2}=(2.7\pm8.5)\times10^{-16}.
\end{align}
These yield the bounds on the theoretical parameters at 95.4\%~CL,
\begin{align}
\om_{\rm BD}\ge3800 \qquad {\rm for~~BD},
\end{align}
and
\begin{align}
\label{psr_con}
\f{\ph}{\Mpl}\le4.4\times10^{-8} \qquad {\rm for~~SMG}.
\end{align}

LLR test of the Nordtvedt effect and a variation in the gravitational constant are given by $\dot{G}/G=(-0.7\pm3.8)\times10^{-13}~{\rm yr}^{-1}$ and $\eta_{\rm N}=(-0.6\pm5.2)\times10^{-4}$ at 68.3\% CL in the literature \cite{Hofmann:2010aa}.
Using the LLR constraint on $\eta_{\rm N}$, we obtain the two following bounds at 95.4\%~CL,
\begin{align}
\om_{\rm BD}\ge1000 \qquad {\rm for~~BD},
\end{align}
and
\begin{align}
\label{LLR_SMG_phi}
\f{\ph}{\Mpl}\le7.8\times10^{-15} \qquad {\rm for~~SMG},
\end{align}
which improves the previous best constraint on $\ph$.
The above results are also shown by the yellow region in Figure \ref{fig_psr_llr_BD_SMG}.
Comparing the pulsar and LLR constraints, we can see that the pulsar constraint is more stringent than the LLR constraint for BD gravity, but for SMG the LLR constraint is around seven order of magnitude more stringent than the pulsar constraint.
This is because in SMG the binary pulsar is a strong screening system relative to the Solar system.

For BD gravity, combining the LLR constraint on $\dot{G}$ and the pulsar constraint on $\dot{P}_b^{\rm Exc}$ from PSR J1738$+$0333, we obtain a better bound of $\om_{\rm BD}\ge13000$ (95.4\% CL), which is still weaker than the limit 40000 from the Cassini experiment \cite{Bertotti:2003aa, Perivolaropoulos:2010aa}.  
Using the parametrized post-Newtonian (PPN) parameter $\gamma=(\om_{\rm BD}+1)/(\om_{\rm BD}+2)$ in BD \cite{Will:1993aa,Will:2014aa}, and combining the Cassini constraint on $\gamma^{\rm obs}=1+(2.1\pm2.3)\times10^{-5}$ \cite{Bertotti:2003aa} and the pulsar constraint on $\dot{P}_b^{\rm Exc}$ from PSR J1738$+$0333, we obtain a better bound of $\dot{G}/G=(-0.2\pm2.3)\times10^{-12}~{\rm yr}^{-1}$ (68.3\% CL) relative to the constraint from two pulsars \eqref{BD_G_W}.

For $G\propto1/\phi_0$ in BD \cite{Will:1993aa,Will:2014aa} and $G\propto1+\ph^2/(2\Mpl^2\Phi_1\Phi_2)$ in SMG \cite{Zhang:2017aa}, the constraint on $\dot{G}/G$ can be translated directly to a bound on the time variation of the scalar background. 
Using the LLR constraint on $\dot{G}/G$, we obtain the two following bounds at 68.3\%~CL,
\begin{align}
\f{\dot{\phi_0}}{\phi_0}=(0.7\pm3.8)\times10^{-13}~{\rm yr}^{-1} \qquad {\rm for~~BD},
\end{align}
and
\begin{align}
\f{\dot{\phi}_{\scriptscriptstyle\rm VEV}}{\ph}=(0.06\pm4.8)\times10^{-4}~{\rm yr}^{-1} \qquad {\rm for~~SMG}.
\end{align}

Note that, we would like to emphasize that above all results about SMG are generically applicable to all SMG theories, which includes chameleon \cite{Khoury:2004aa}, symmetron \cite{Hinterbichler:2010es}, dilaton \cite{Damour:1994ab}, and $f(R)$ \cite{Sotiriou:2010aa,De-Felice:2010aa} theories. 
In the next section we consider the chameleon model as an example.

\section{Application to Chameleon}\label{section4}
The most important one of the SMG theories is the chameleon model, introduced as a screening mechanism by Khoury and Weltman \cite{Khoury:2004aa}. 
The chameleon operates an environment-dependent scalar field, which can acquire a large mass in dense regions to suppress the fifth force. 
The original model has already been ruled out by the combined constraints from the Solar system and cosmology \cite{Hees:2012aa, Zhang:2016aa}. 
However, the idea of chameleon can be resurrected by an exponential potential and an exponential coupling function \cite{Brax:2004aa},
\begin{align}\label{Chameleon_V_A}
V(\phi)=\Lambda^4\exp\Big(\frac{\Lambda^{\alpha}}{\phi^{\alpha}}\Big),\qquad
A(\phi)=\exp\Big(\frac{\beta\phi}{\Mpl}\Big), 
\end{align}
where $\alpha$, $\beta$, and $\Lambda$ are a dimensionless constant index, a dimensionless coupling constant between chameleon and matter, and the energy scale of the theory, respectively.

\begin{figure}[htbp]
\centering
\includegraphics[width=8cm, height=6cm]{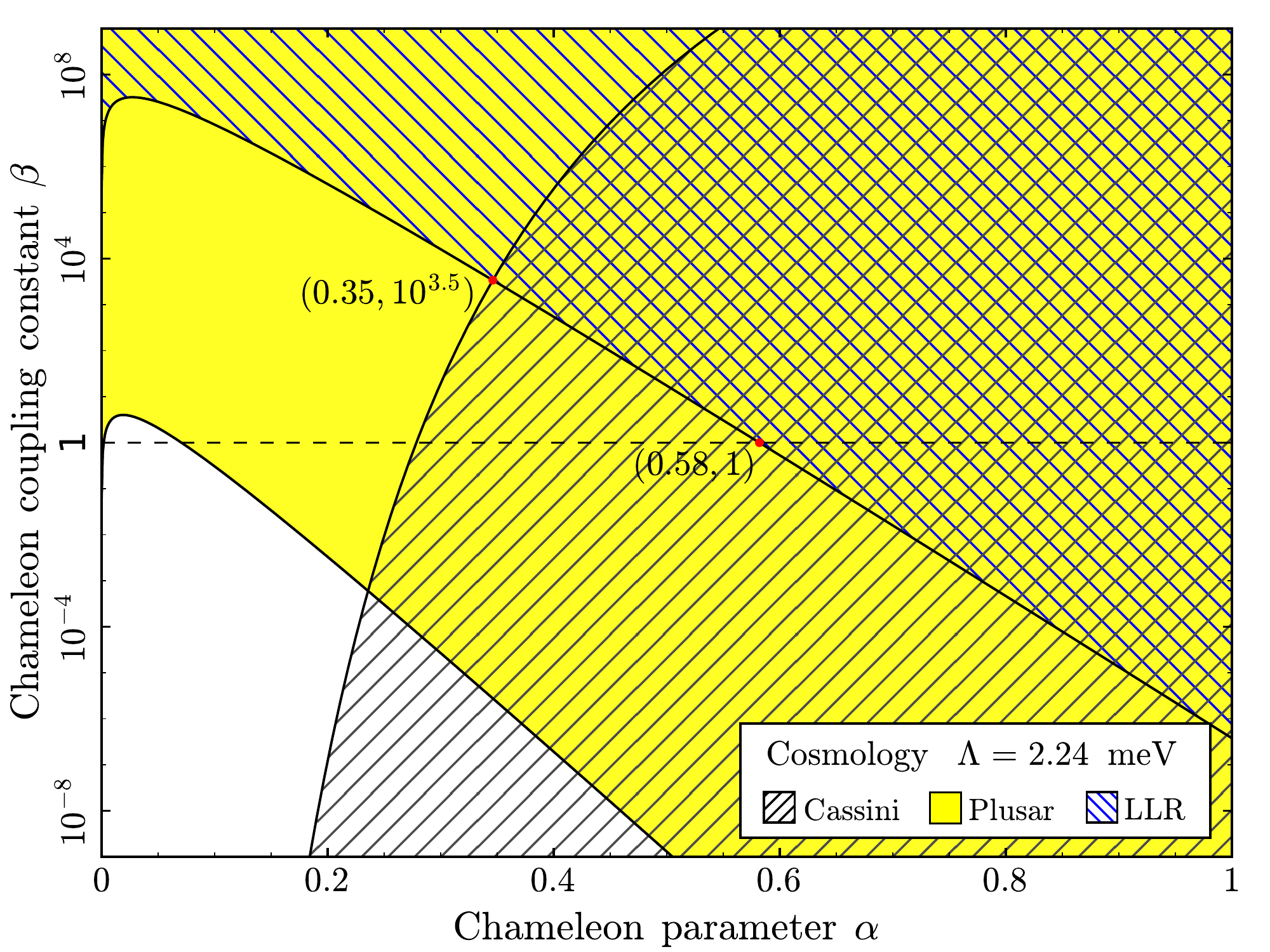}
\caption{
In the parameter space of exponential chameleon model, the gray shaded region, the blue shaded region, and the yellow region are allowed by Cassini probe, LLR measurements, and pulsar observations from PSRs J1738$+$0333 and J0348$+$0432, respectively. 
The overlap region yields a lower bound of $\alpha\ge0.35$.}
\label{fig_psr_llr_cas_Cha}
\end{figure}

The cosmological constraints require that the energy scale $\Lambda$ is close to the dark energy scale of $2.24\times10^{-3}~\rm eV$ \cite{Zhang:2016aa,Hamilton:2015aa}. 
The chameleon VEV is given by $\ph=[{{\alpha} \Mpl \Lambda^{4+\alpha}}{(\beta\rho_{b}})^{-1}]^{\f{1}{\alpha+1}}$ \citep{Zhang:2016aa}, where $\rho_{b}$ is the background matter density.
Using this and the best constraint on $\ph$ \eqref{LLR_SMG_phi} from LLR, we present the allowed region of the parameter space $(\alpha,\,\beta)$, marked as the blue shaded region in Figure \ref{fig_psr_llr_cas_Cha}.
The pulsar constraint \eqref{psr_con} from PSRs J1738$+$0333 and J0348$+$0432 is illustrated in Figure \ref{fig_psr_llr_cas_Cha} by the yellow region.
In addition, the PPN parameter is given by $\gamma=1-2\beta\ph/(M_\text{Pl}\Phi)$ \cite{Zhang:2016aa}, from the Cassini constraint $\left|\gamma_{\rm obs}-1\right|\le2.3\times10^{-5}$ \cite{Bertotti:2003aa}, the allowed region is shown by the gray shaded region in Figure \ref{fig_psr_llr_cas_Cha}.
We can see from this figure that $\alpha\ge0.58$ if the chameleon is weakly coupled to matter ($\beta\le1$).
Combining the LLR and Cassini constraints gives a stringent bound of $\alpha\ge0.35$, which improves slightly on the previously published limit of $0.2$ \cite{Zhang:2017aa}.

\section{Conclusion}\label{section5}
As a simple generalization of GR, SMG and BD gravities are scalar-tensor theories of gravity with and without screening mechanisms, respectively.
In this paper, we investigated the tests of the existence of dipole radiation and time-varying $G$ and the Nordtvedt effect in BD and SMG theories using Cassini experiment, LLR measurements, and pulsar observations from PSRs J1738$+$0333 and J0348$+$0432, respectively.
We derived new constraints on dipole radiation and time-varying $G$ in these theories by combining these tests.
We found that screening mechanism has significant impact on the model constraints.
For BD gravity, the strongest, second, and weakest constraints are from Cassini experiment, pulsar observations, and LLR measurements, respectively.
Because of the existence of screening mechanism in SMG, the LLR constraint on SMG is several order of magnitude more stringent than the pulsar observations, which also improves the previous best constraint on the scalar background.
The LLR constraint on $\dot{G}$ is also more stringent than the pulsar observations.
We translated the constraint on $\dot{G}$ into the bounds on the time evolution of the scalar background in these theories, which differ by several orders of magnitude.

As an application, we applied our results to the exponential chameleon model. 
We derived the combined constraints on the model parameters from Cassini, LLR, and pulsar tests, and obtained a new bound on the chameleon parameter of $\alpha\ge0.35$. 
Finally, we would like to emphasize that the results of all tests show good agreement with GR and all the previous works and give more stringent constraints on the deviations from GR.

\begin{acknowledgments}
This work is supported by NSFC No. 11603020, 11633001, 11173021, 11322324, 11653002, 11421303, project of Knowledge Innovation Program of Chinese Academy of Science, the Fundamental Research Funds for the Central Universities and the Strategic Priority Research Program of the Chinese Academy of Sciences Grant No. XDB23010200.
\end{acknowledgments}

\bibliography{BD_SMG.bbl}

\end{document}